\documentclass[twocolumn,twoside,slac]{revtex4}
\usepackage{graphicx}
\usepackage{fancyhdr}
\begin{document}

\title{Securing a HENP Computing Facility}

%

\author{S. Misawa, O. Rind, T. Throwe}
\affiliation{Brookhaven National Laboratory, Upton, NY 11973, USA}

\begin{abstract}
Traditionally, HENP computing facilities have been open facilities
that are accessed in many different ways by users that are both
internal and external to the facility.  However, the need to protect
the facility from cybersecurity threats has made it difficult to
maintain the openness of the facility to off-site and on-site users.

In this paper, we discuss the strategy we have used and the
architecture we have developed and deployed to increase the security
the US ATLAS and RHIC Computing Facilities, while trying to maintain
the openness and accessibility that our user community has come to
expect.  Included in this discussion are the tools that we have used
and the operational experience we have had with the deployed
architecture.
\end{abstract}

\maketitle

\thispagestyle{fancy}


\section{Background}
The RHIC Computing Facility (RCF) at the Brookhaven National
Laboratory (BNL) is the primary data archiving and processing facility
for the Relativistic Heavy Ion Collider (RHIC) at BNL.  The US ATLAS
Computing Facility at BNL is the US Tier 1 computing facility for the
ATLAS experiment at the Large Hadron Collider (LHC) at CERN.  As major
computing centers, these facilities provide a myriad of services to
users that are on-site at BNL and off-site at collaborating
institutions.  These services include the archive storage of raw and
analyzed data from the experiments, interactive and batch data
processing services and general interactive login facilities.
Traditionally, such facilities have been completely open to the
Internet and have provided services, e.g. FTP (File Transfer
Protocol), telnet, IMAP (Internet Message Access Protocol), that were
not designed with cybersecurity in mind.

An intrusion into either facility by a malicious intruder can result
in damages from which recovery can be extremely costly in time and
money.  Some of the difficulty is associated with the enormous
facility ``state'' represented by the 100's of terabytes of on line
disk storage and the 1000's of compute nodes at the facility. However,
this is not the limits of the damage. Since the facility is part of
the data collection process, disruption of the facility can also
result in the loss of data taking ability at each experiment. This
loss is extremely costly since operating an accelerator is an
expensive undertaking. Finally, because the facility is the sole
archive of a substantial fraction of raw and analyzed data created by
the experiments, loss of these archives from actions by malicious
intruders can result in irreparable damage to on going scientific
research.

Although many other computing facilities are at risk from
cybersecurity incidents, two properties make the RCF and US ATLAS
computing facilities high profile targets. First, the vast amounts of
computing resources make the facilities appealing targets to
intruders. Second, the location of the facility within a US
government, Department of Energy facility, with its associated .gov
top level domain makes these centers choice targets for politically
motivate malicious intruders.

With the proliferation of Internet connectivity outside of the
academic and research communities, the open data center architecture
needs to be modified to mitigate the increased risk of cybersecurity
threats.

\section{Securing the Facility}
The process of improving the security of the facilities was a
multi-stage process. The first step involved the identification of
assets and services at the facilities. In the second step, priorities
were assigned to the identified asset or service with respect to
importance. The third step was identifying the potential threats to
the assets or services. The fourth step was modifying the network
topology of the facility to mitigate the potential threats. The fifth
step was replacing insecure services with more secure services. The
sixth step was the modification of user and administrator behavior to
increase the security of the facility. The seventh step was an
assessment of the effectiveness of the instituted changes. The final
step was the iteration of the proceeding seven steps to continue to
improve the security of the facility.

\subsection{Assets and Services}
There are a number of assets and services at the RCF and US ATLAS
computing facilities with differing security needs. The main data
store, the High Performance Storage System (HPSS)~\cite{hpss-ref},
contains the raw and analyzed data that is generated from the
experiments. The protection of the stored data and the operation of
the service are paramount to the facility. Without it, the other
components of the facility are inconsequential. The service provide by
HPSS is data storage and retrieval via an FTP-like mechanism. The
operation of the system is effectively independent of the other assets
and services in the facility.

After HPSS, the next service of importance at the facility is the
authentication and authorization infrastructure, consisting of a
cluster of Network Information System (NIS) servers. Without account
information, the compute farm and the NFS (Network File System)
storage are unusable. As a result, the protection of this information
is crucial.

At approximately equivalent levels of importance are the NFS servers
and the Linux based computation farm. The NFS storage consists of
approximately 150TB of disk space and 20 NFS servers. Although,
extremely disruptive, loss of NFS data or service is not
catastrophic. Lost unprocessed data can be restored from HPSS and lost
processed data can be recovered by reprocessing the raw data. However,
depending on the details of the data loss, the recovery process may be
short and simple or long and complex. The Linux based computation farm
consists of approximately 1100 dual CPU systems and 130TB of local
storage. Loss of these systems would halt data processing and loss of
data stored on the local disks would require reloading of data from
the data source or the recreation of data by reprocessing. As with the
NFS servers, the type of loss would dictate the cost of recovery.

The wide area accessible Andrew File System (AFS)~\cite{openafs-ref}
storage is at the next level of importance at the facility. Loss of
this service would be inconvenient for data processing on
site. Recovery of data corrupted AFS storage would entail restoration
from backups, resulting in a maximum loss of about 24 hours of
work. Loss of AFS service would definitely cause major disruption to
work at off site locations. For on site users, loss of AFS service
can, if necessary, be supplanted with NFS service.

The remaining services provided by the RCF and US ATLAS facilities
include web service, Samba service, and email service. Although
important for the users, these services are not coupled to the higher
priority mission of the facilities, which is to provide storage
and computational resources to the individual experiments.

With the assets and services at the facilities identified, an
assessment of the security threats is necessary to determine the
changes that need to be made to increase the security of the
facilities.

\subsection{Assessing the Threats}
The threats to computing resources at the RCF and US ATLAS computing
facilities fall into two classes, direct network assaults and
compromised accounts. Examples of direct network assaults include
direct attacks on network service daemons (e.g., buffer overflow
exploits, protocol weaknesses), hijacking of network connections
(e.g., replay attacks, forged packets), denial of service (DoS),
distributed DoS, and web based attacks (both client and server). This
class of attacks do not require login access on a facility
system. Examples of compromised accounts include stolen or cracked
passwords, social engineered access, dead accounts, and malicious
insiders. This class of attacks involve gaining access to a login
account at the facility from which exploits are launched. An
interesting distinction between these two classes of assaults is the
mitigation of network attacks can be more easily accomplished with
technology and architecture changes. On the other hand, behavioral
changes on the part of the system administrators or users are usually
required to fix attacks in the compromised accounts class.

\subsection{Facility Architecture}

A modification of the facility architecture is the simpler of the two
classes of changes that were made at the RCF and US ATLAS
facilities. (The harder change being the modification of user and
administrator behavior.) The two main goals of the architectural
modifications were compartmentalization and isolation. Where possible,
unrelated assets and services were separated so that security breaches
in one location did result in immediate security breaches in another
location. Also, where possible, network accessability to assets and
services was minimized.

The core facility, consisting of the HPSS servers, NFS servers, AFS
servers, NIS servers and Linux computation farm, were placed behind a
firewall with a ``default deny'' firewall rule. With the exception of
the AFS servers and AFS service, there is virtually no direct network
access from the outside to the core facility. In this configuration,
the experiment counting houses, where raw experiment data is
generated, are placed within the firewall.

For scalability and availability, the facility firewall was a
multi-component firewall, consisting of a Cisco PIX firewall,
augmented by ssh~\cite{openssh-ref}, secure file transfer and
Samba~\cite{samba-ref} (SMB) gateways. The gateways were implemented
using general purpose computers running host based firewalls
(ip-filter~\cite{ip-filter-ref} or iptables~\cite{iptables-ref}) on
stripped down and hardened operating system installations. Each
gateway system is equipped with two network interfaces (NIC), allowing
them to bypass the PIX firewall, thus providing some scalability and
availability. This dual NIC configuration is what dictated the need
for the host based firewalls.  The separation of the ssh and data
transfer gateways into separate systems simplifies the hardening and
maintenance of each system. It also isolates interactive login traffic
with its low latency requirements from data transfers which are
relatively insensitive to latency and typically result in high system
load.

In conjunction with the introduction of the gateways was the
elimination of all protocols with clear text passwords. FTP data
transfers were replaced with scp~\cite{openssh-ref},
bbftp~\cite{bbftp-ref}, and ssh protected ftp. (The latter being
normal FTP with the FTP control channel being sent through an
encrypted ssh tunnel.) Interactive logins via telnet and rsh were
replaced with encrypted logins via ssh. This latter change providing
the additional benefit of protecting X~\cite{x-ref} window traffic.

With the new core facility configuration, triage scenarios are
possible with respect to intrusions. The first level would involve
disconnecting the facility from the Internet, thus allowing on site
user to work and on going operations to continue. The second level
involves disconnection from the on site network. This would allow on
going operations to continue, including data taking, but would cut off
on site and off site users. The third level would involve the shutdown
of all servers and services except for HPSS, thus halting all
operation except data taking.

Located both outside and inside the facility firewall are multiple web
servers. The services provided and resources used by the web servers
are carefully distributed to the appropriate web server. Located
outside the firewall is a non authenticating web server, serving both
static and dynamic (aka CGI scripts) pages. This web server is
completely stand alone, requiring no services or resources from
systems inside the firewall. Two web servers are located inside the
firewall, an authenticating web server and a non authenticating web
server. The authenticating web server serves both static and dynamic
web pages, but requires users to authenticate to the server before
access is granted to the pages. (The authentication occurs over an SSL
protected connection~\cite{openssl-ref}.) The assumption is that
authenticated users, i.e. legitimate facility users, are not malicious
and will not attempt to exploit weakness that may be present in the
dynamic web pages. This web server is located behind the facility
firewall because of its dependence on a core facility service. The
second web server located inside the firewall is a non authenticating
web server that serves only static user pages. Since it depends on
access to user home directories, it resides behind the firewall. To
eliminate the problems associated with CGI scripts and other dynamic
content, only static pages are served.

The final component of the new facility architecture was the
separation of email services from the core facility. A standalone SMTP
and secure IMAP server~\cite{imap-ref} was installed outside of the
facility firewall. The email server utilizes a separate password
database that is independent of the password database used for
interactive access to the core facility. This separate database can
potentially isolate the core facility from email account
compromises. In addition, the use of SSL protected IMAP (replacing
standard IMAP) protects email passwords while they traverse the
network.

\subsection{Human Factors}
Architectural changes to the RCF and US ATLAS facilities combined with
technology choices significantly enhances the security of the
facilities. However, a substantial amount of the security is dependent
on the actions and behavior of users and administrators. As an
example, benefits of encrypted ssh connections to the facility are
negated if users explicitly set the X~window DISPLAY variable to
bypass the encrypted tunnel, or use telnet to log into the system from
which he/she runs the ssh client. Similarly, system administrators can
destroy site security if they are ``social engineered'' into providing
account access to unauthenticated or unauthorized individuals.

Change user and administrator behavior primarily consists of education
to make people aware of the security issues and to provide information
on secure ways of doing tasks. However, education is not exceedingly
effective. There is a distinct difference between awareness and
understanding and theory and practice. Some behavior was enforced, for
example disabling of telnet and proactive password checking with
npasswd~\cite{npasswd-ref}. Unfortunately, most behavior cannot be
enforced.

\section{Operational Assessment}
Operational experience that has been collected with the new facility
architecture in place provides information about what is working and
what is not. As was expected, migration to the new configuration was a
long and drawn out process, much akin to changing the direction of an
oil supertanker. New tools and configurations needed to be tested,
documented and put into production. In addition, generous lead time
was required so that users could prepare for the system changes. At
this point, the facility has accomplished the first iteration of the
facility hardening process outlined at the beginning of the previous
section.

The critical components of the new facility are in the multicomponent
firewall, since they represent the interface to the facility for the
users. The most frequently used component in the firewall are the ssh
gateways. Implemented using four 650MHz Intel Celeron PCs, they are
each able to handle up to 80 interactive users running xterms. However,
a single graphics intensive application can saturate the CPU,
resulting in longer latencies for users. Additional problems have been
encountered with scp data transfers on the ssh gateways. As with
graphics intensive applications, a single scp session can saturate the
CPU resulting in responsiveness issues for other users. The ssh
gateways have subsequently been upgraded to 1.7GHz Pentium~4 systems,
providing substantially improved performance.

The FTP gateway, currently a dual 800MHz Pentium III system with dual
gigabit ethernet interfaces, has been a little problematic. Getting
users to use the FTP gateway was the first problem that was
encountered. Over time, this has become less of an issue. An on going
problem is the use of sftp and scp for bulk data transfers. As on the
ssh gateways, a single sftp/scp transfer can saturate a single
CPU. For a handful of power users, bbftp has worked well for bulk data
transfers since only the control channel is encrypted. However, its
non FTP-like user interface and its relative obscurity compared to scp
and FTP are a barrier to wide spread adoption by users. Additional
problems are associated with the use of NFS mounted disks as a source
and destination for data files on the gateway. The performance of the
NFS disk is a bottleneck for high speed transfers of individual
files. The installation of local disk is an option but creates
problems with disk space management. To alleviate some of the
performance issues, the FTP gateway will be replaced with two dual
2.4GHz Pentium~4 systems.

For all components of the firewall, configuration and maintenance is
problematic. The maintenance of the rules in the PIX firewall to
handle the addition, removal, and movement of services is a
significant undertaking. Judging the security of new service is
particularly problematic. In addition, some services are difficult to
protect with a firewall. With the host-based firewalls, the
configuration of the rules can be tricky and can result in unexpected
problems.

Finally, significant work is still necessary on audit trails and
intrusion detection on the various firewall components.

\section{Future Work}

Operational experience has revealed areas where modifications can be
made to improve site security. In addition, the Internet and
cybersecurity threats are constantly changing, as a result, defenses
need to be upgraded to handle them. Since most of the security
features in the new facility architectures are designed to handle the
threats circa the late 1990's, work needs to be done to deal with new
types of threats.

One area of work is the additional hardening of the ``edge'' systems,
i.e., those systems that are directly accessible from the
Internet. Better understanding of the issues and technologies makes it
possible to better protect the systems and the facility. Examples of
these changes include better firewall rules, updated software, and
additional re-architecting of services to improve security and
manageability.

With the proliferation of sophisticated web services, the protection
of web clients from malicious web servers is also an idea to be
considered. On a similar vein, proactive institution of firewall rules
to mitigate the risk of the facility being a launch site of a network
attack is also worth considering.

\subsection{Grid Services}

Looming on the horizon are issues presented by Grid~\cite{globus-ref}
technology. With the Grid comes new types of security concerns. One
promise of Grid technologies is the availability of global computation
resources from virtually anywhere in the world. However, this also
adds a new scale to the ramification of security breaches, e.g.,
access to global computational resources to launch distributed DoS
attacks and the destruction or disabling of resources on a global
scale. With the Grid, responsibilities, facilities and authorities
become geographically and organizationally distributed. Command,
control and communication become much more difficult.

Operationally, the phasing in of incomplete or untested Grid services
into a production facility has unknown risks. Additionally, the skill
sets and experience need to manage these new services do not exist and
will likely result in operational errors which may have dire security
consequences.

\section{Conclusions}

With a modification of the architecture of the RCF and US Atlas Tier 1
computing facilities, the replacement of insecure services with more
secure services, and the education of both facility users and
administrators, substantial gains in facility security have been
obtained. However, security is an on going process, not an
endpoint. The introduction of new requirements and services are making
security harder and security breaches potentially more
disastrous. Maintaining the integrity of the current security profile
while growing the facility and accommodating new requirements and
services will make security an increasingly difficult task.

\begin{acknowledgments}
The authors wish to thank the five RHIC experiments, the US ATLAS
group, the BNL Physics department, and the Network group in the
Information Technology Division at BNL.

\end{acknowledgments}


\end{document}